\documentclass[twocolumn, graphicx]{revtex4-1}

\usepackage{graphicx}
\usepackage{textcomp}
\usepackage{hyperref}
\usepackage{color}

\draft 

\begin{document}

\title{Thermal release tape-assisted semiconductor membrane transfer process for hybrid photonic devices embedding quantum emitters} 

\author{Cori Haws}
\affiliation{James Watt School of Engineering, Electronics \& Nanoscale Engineering Division, University of Glasgow, Glasgow G12 8LT, United Kingdom}
\affiliation{Physical Measurement Laboratory, National Institute of Standards and Technology, 100 Bureau Drive Gaithersburg, MD 20899, USA}

\author{Biswarup Guha}
\affiliation{Physical Measurement Laboratory, National Institute of Standards and Technology, 100 Bureau Drive Gaithersburg, MD 20899, USA}
\affiliation{Joint Quantum Institute, NIST/University of Maryland, College Park, MD 20742, USA}

\author{Edgar Perez}
\affiliation{Physical Measurement Laboratory, National Institute of Standards and Technology, 100 Bureau Drive Gaithersburg, MD 20899, USA}
\affiliation{Joint Quantum Institute, NIST/University of Maryland, College Park, MD 20742, USA}

\author{Marcelo Davanco}
\affiliation{Physical Measurement Laboratory, National Institute of Standards and Technology, 100 Bureau Drive Gaithersburg, MD 20899, USA}

\author{Jin Dong Song}
\affiliation{Center for Opto-Electronic Materials and Devices Research, Korea Institute of Science and Technology, Seoul 136-791, South Korea}

\author{Kartik Srinivasan}
\affiliation{ Physical Measurement Laboratory, National Institute of Standards and Technology, 100 Bureau Drive Gaithersburg, MD 20899, USA}
\affiliation{Joint Quantum Institute, NIST/University of Maryland, College Park, MD 20742, USA}

\author{Luca Sapienza\\
E-mail address: \href{mailto:luca.sapienza@glasgow.ac.uk}{luca.sapienza@glasgow.ac.uk}\\
Website: \href{https://sites.google.com/view/integrated-quantum}{https://sites.google.com/view/integrated-quantum}}
\affiliation{ James Watt School of Engineering, Electronics \& Nanoscale Engineering Division, University of Glasgow, Glasgow G12 8LT, United Kingdom}
\affiliation{Physical Measurement Laboratory, National Institute of Standards and Technology, 100 Bureau Drive Gaithersburg, MD 20899, USA}

\date{\today}

\begin{abstract}

Being able to combine different materials allows taking advantage of different properties and device engineering that cannot be found or exploited within a single material system. In the realm of quantum nano-photonics, one might want to increase the device functionalities by, for instance, combining efficient classical and quantum light emission available in III-V semiconductors, low-loss light propagation accessible in silicon-based materials, fast electro-optical properties of lithium niobate and broadband reflectors and/or buried metallic contacts for local electric field application or electrical injection of emitters. However, combining different materials on a single wafer is challenging and may result in low reproducibility and/or low yield. For instance, direct epitaxial growth requires crystal lattice matching for producing of defect-free films, and wafer bonding requires considerable and costly process development for high bond strength and yield. We propose a transfer printing technique based on the removal of arrays of free-standing membranes and their deposition onto a host material using a thermal release adhesive tape-assisted process. This approach is versatile, in that it poses limited restrictions on the transferred and host materials.
In particular, we transfer 190\,nm-thick GaAs membranes, with dimensions up to about 260$\,\mu$m\,$\times$\,80$\,\mu$m, containing InAs quantum dots, onto a gold substrate. We show that the presence of a back reflector combined with the etching of micro-pillars significantly increases the extraction efficiency of quantum light, reaching photon fluxes, from a single quantum dot line, exceeding 8\,$\times$\,10$^5$ photons per second, which is four times higher than the highest count rates measured, on the same chip, from emitters outside the pillars. Given the versatility and the ease of the process, this technique opens the path to the realisation of hybrid quantum and nano-photonic devices that can combine virtually any material that can be undercut to realise free-standing membranes that are then transferred onto any host substrate, without specific compatibility issues and/or requirements.
\newpage
\end{abstract}


\maketitle 

The development of hybrid integrated photonic devices has attracted considerable interest, for instance in the field of silicon photonics, in order to realise III-V lasers on silicon \cite{Huiyun} and light-emitting diodes on flexible substrates \cite{LEDs}. More recently, hybrid approaches have been investigated for quantum photonic applications \cite{rev,Edo}. There are several different approaches for realizing hybrid integration, including wafer bonding \cite{us} and pick-and-place of nanophotonic devices \cite{Tokyo}. These approaches have led to the successful demonstration of low-loss guidance of quantum light \cite{Tokyo}, cavity quantum electro-dynamics on a III-V/silicon platform \cite{us}, on-chip interference from quantum dots in silicon nanowires \cite{Val} and single-photon indistinguishability in hybrid devices \cite{Marcelo}. An alternative route is represented by transfer printing, where devices or patches of material are transferred from one substrate to another, an approach that has been widely exploited in electronics \cite{memb, memb2, memb3, memb4}, which has resulted in the realization of detectors and flexible devices \cite{flex1, flex2, flex3, flex4, flex5}, and has also more recently been applied to quantum nanophotonics \cite{Tokyo,QP2,QP3, Dirk, Dirk2, Brat}.

Here, we propose the use of a membrane-transfer technique for an easy integration of different (and potentially multiple) materials on the same wafer. Our technique relies on a thermal-release tape that is commonly used to exfoliate two-dimensional materials \cite{2D, 2Db} and we used it to integrate active quantum emitters on a host substrate. Compared to polydimethylsiloxane (PDMS)-based processes, our approach replaces the patterning and kinetic control of adhesion required when utilizing elastic stamps with membrane release that is controlled by the tape's temperature. Given the relatively large area of material deposited, compared to the footprint of standard micro- and nano-scale devices, hundreds of devices can then be nano-fabricated on the hybrid chip, for instance to control light emission and propagation.

\begin{figure*}[ht!]
\centering
\includegraphics[width=1\linewidth]{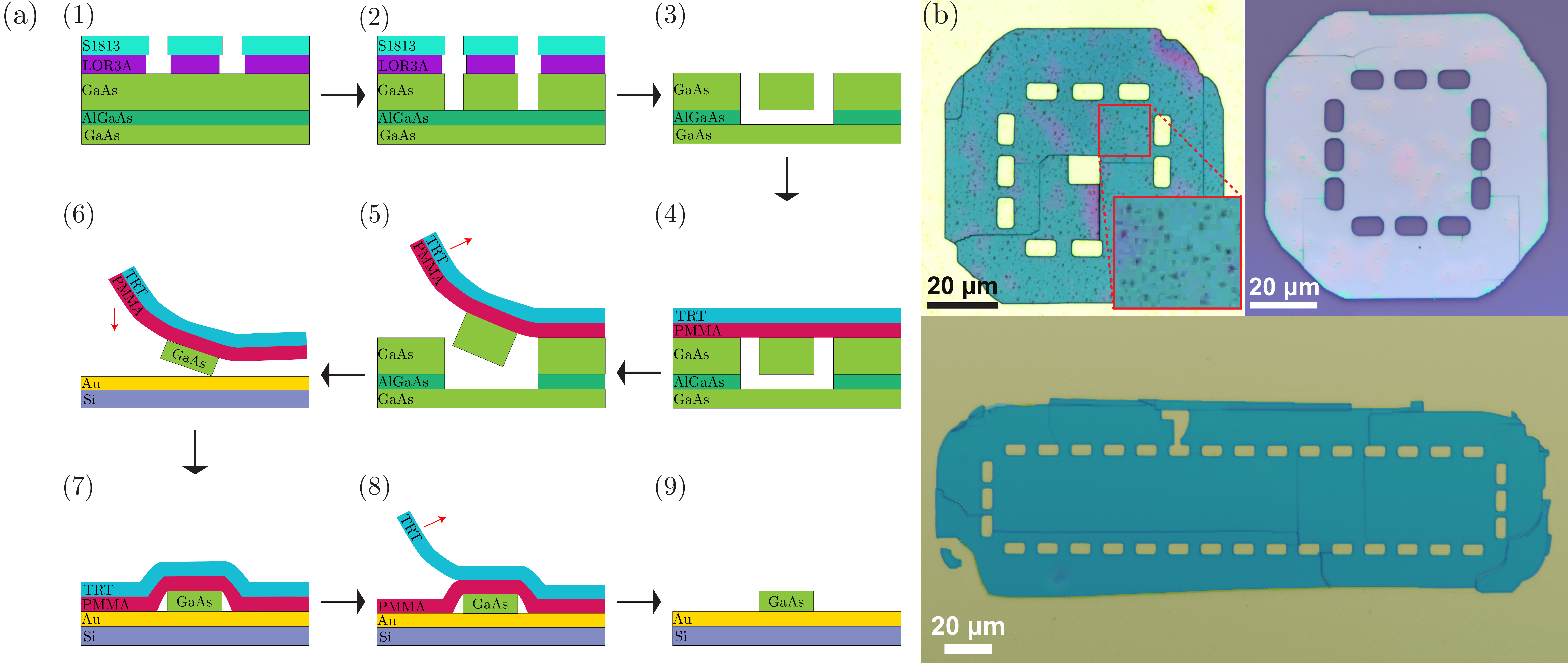}
\caption{(a) Schematic (not to scale) of the membrane transfer process used to realise hybrid devices. (1) A direct laser write is used to write a set of of apertures (the holes appear as `dashed lines' that define the perimeter of the membranes that will be transferred) on a resist layer. (2) The apertures are then dry etched through the GaAs top layer to reach the AlGaAs sacrificial layer. (3) A hydrofluoridic acid wet etch is then used to undercut the membranes. (4) Resist (PMMA) is then deposited on the chip to protect the GaAs from impurities during the transfer process that is carried out by placing a thermal release tape (TRT) on the sample surface. (5) The tape, that has now the membranes attached, is manually removed (6) and placed on a host substrate (7). (8) The tape is heated so that it loses adhesion and leaves the membranes on the host substrate where they are held in place by van der Waals forces. (9) The resist layer that protected the GaAs membrane is then chemically removed with a solvent. (b) Images of membranes transferred using the technique described in (a). The top left (right) panel shows a membrane transferred without (with) the resist protective layer (step 4 in panel (a)): tape residues are visible on the surface of the membrane on the left, while the surface of the membrane on the right appears clean.}
\end{figure*}

The transfer process that we have developed is schematically illustrated in Fig.\,1a. A direct laser write maskless aligner (equipped with a laser emitting at 405\,nm) is used to write a series of apertures (the holes appear as `dashed lines' that define the perimeter of the membranes that will be transferred) on a photoresist film ($\approx$1.6\,$\mu$m thick). These apertures are then dry-etched through the GaAs top layer to reach the AlGaAs sacrificial layer. The residual resist is chemically removed with a solvent, and a hydrofluoridic acid wet etch (49\,$\%$ concentration by mass in water) is used to undercut the membranes. The samples are then cleaned in de-ionised water and isopropyl alcohol and dried by heating to about 120\,$^{\circ}$C. Resist is then deposited, via standard spin coating, on the chip ($\approx$150\,nm of PMMA 495A4 \cite{disc}
) to protect the GaAs from impurities during the transfer process that is carried out by slowly, manually placing a thermal release tape (Graphene Supermarket GTT-DS \cite{disc}
) on the sample surface (see Fig.\,1b). The tape, that now has the membranes attached, is then manually removed and placed on a host substrate. The sample is then slowly heated up to about 100\,$^{\circ}$C so that it loses adhesion and leaves the membranes on the host substrate where they are held in place, by van der Waals forces. Finally, the resist layer that protected the GaAs membrane is chemically removed with a solvent. It is worth noting that without the resist protective layer, the adhesive tape would leave residues on the sample surface, residues which can be difficult to remove and that can affect the optical quality of the fabricated devices. The presence of the PMMA reduces the yield in the pick up process but guarantees cleanliness of the membrane top surface (see Fig.\,1b).\\
Figure\,1b shows examples of membranes of different sizes and shapes that we have transferred using the technique presented above. Arrays of hundreds of membranes can be transferred in one single step, in our case from samples as large as 2\,cm\,$\times$\,2\,cm (larger samples might be possible, but have not been tested), and the yield, defined as the number of transferred membranes that have more than 80\,$\%$ of usable area, is generally around 75\,$\%$.
Other approaches, for instance those based on PDMS as material to pick up the devices \cite{Tokyo}, require fine control on how the polymeric stamp is placed on the sample surface and how it is peeled off to release the transferred material. Our technique, instead, takes advantage of the fact that the tape has very strong adhesion until it is heated above a certain temperature, after which it will lose adhesion and it will release the vast majority of the membranes (more than about 90\,$\%$). A disadvantage is related to the need to protect the sample surface, in our case with a resist layer, which, as discussed above, can reduce the adhesion and therefore overall yield of the transfer process. Compared to wafer-bonding approaches \cite{us}, our transfer process is much more tolerant of surface roughness, eliminates time-consuming substrate removal steps, and materials can be directly bonded together without requiring intermediate matching layers.

\begin{figure*}[t!]
\centering
\includegraphics[width=\linewidth]{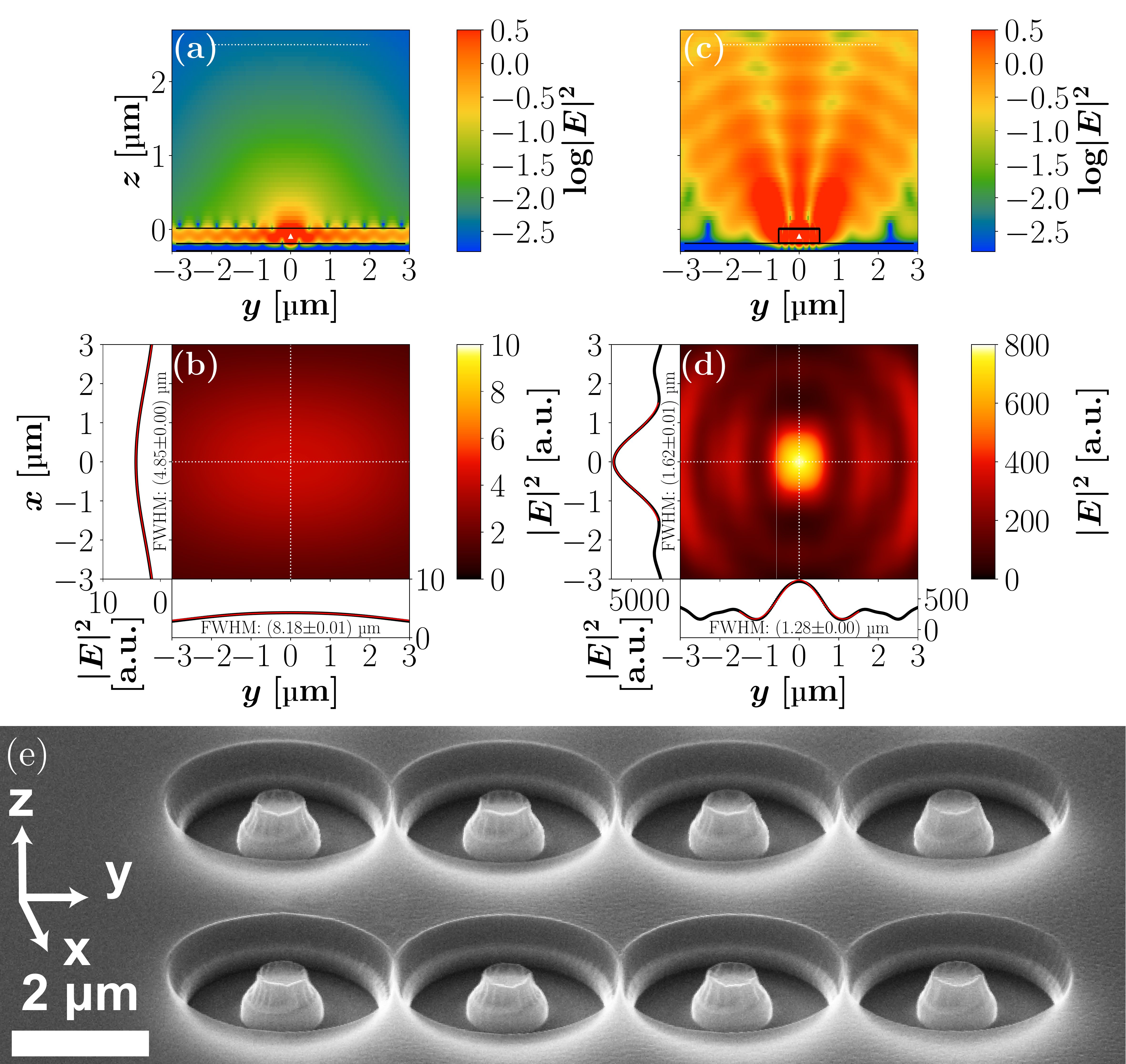}
\caption{(a) Colour plot, obtained by finite-difference time-domain simulations, of the side ($zy$ plane) profile of the squared electric field of a dipole emitter (oriented along the $x$ axis and emitting between 925\,nm and 975\,nm) placed in the middle of a 190\,nm-thick GaAs membrane transferred to a Au layer, and (b) top view ($xy$ plane), measured at a distance of 2.5\,$\mu$m from the sample surface in correspondence to the dotted line shown in panel (a), including linecuts of the far-field profile (black lines) and their Gaussian fits (red lines). The Full-Width Half-Maximum (FWHM) obtained from the fits is indicated in the graphs, with the error obtained from the fitting function (one standard deviation uncertainty due to the fit negligible compared to the values). (c, d) Same as panels (a, b) in the presence of a micro-pillar with diameter of 1\,$\mu$m on the sample surface, centered around the emitting dipole. (e) Scanning electron micrograph image of GaAs micro-pillars (with remnant resist) fabricated via ion milling after deposition, exposure and development of electron-beam resist.}
\end{figure*}

We use this process to transfer 190\,nm-thick GaAs membranes containing, in the centre (95\,nm below the sample surface), a layer of InAs quantum dots, grown by molecular beam epitaxy with a Stranski-Krastanov technique. As discussed, this membrane transfer process allows combining materials that can not be grown with the same techniques and/or in the same chambers. In this case, we deposit the GaAs membrane on a gold-coated Silicon wafer and use the metal as a broadband back reflector. The presence of the back reflector alone (without any other fabricated devices), allows increasing the amount of light collected above the sample by about a factor two, compared to the quantum dot emission as-grown. Figure\,2 shows finite-difference time-domain simulations of the emission of an oscillating dipole within the GaAs membrane under study deposited on a gold layer. Panels (a,b) show the profile of the emitted intensity and panels (c,d) show how, by etching a micro-pillar of 1\,$\mu$m of diameter, centered around the emitter, the light otherwise trapped within the membrane can now be guided vertically out of the chip (and towards an external collection objective or optical fiber). 
Figure\,2e shows examples of micro-pillars fabricated by electron-beam lithography, followed by argon ion milling, on the sample under study. The ion milling is used to avoid contamination of dry etcher chambers, due to the presence of the gold layer underneath the GaAs membrane, and a secondary ion mass spectrometry endpoint detector is used to stop the etching once the gold layer is reached.

\begin{figure*}[t!]
\centering
\includegraphics[width=1\linewidth]{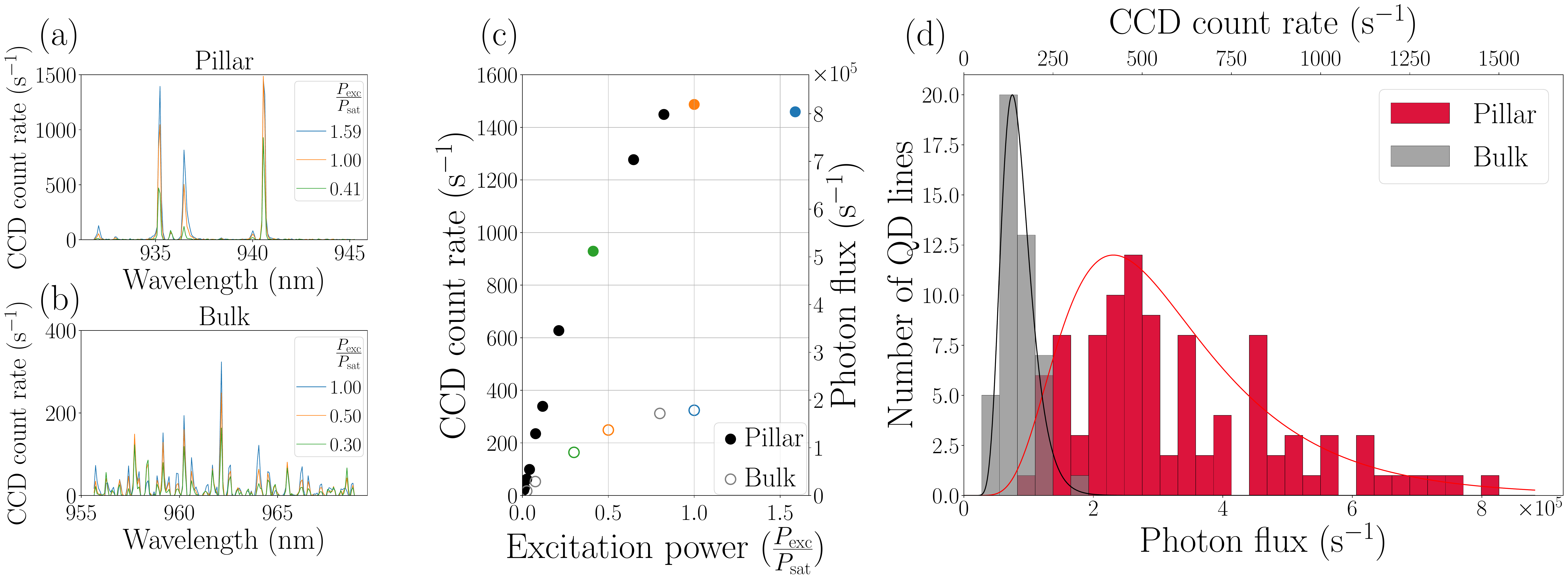}
\caption{(a,b) Example of photoluminescence spectra, collected using an objective with a numerical aperture  of 0.9, at a temperature of about 4\,K, on a Charge-Coupled Device (CCD) placed at the exit of a 0.5\,m-long spectrometer, of quantum dots placed within (a) and outside (b) micro-pillars respectively. (c) Photoluminescence intensity of the emission lines with wavelength of 940.55\,nm in panel (a) and 962.14\,nm in panel (b) (colour coded accordingly), as a function of excitation power of a continuous wave 780\,nm laser (P$_{exc}$), normalised to the power giving the highest emission count rate for a specific emission line (P$_{sat}$\, of \,170\,$\mu$W and 400\,$\mu$W, respectively). The photon flux corresponds to the number of single photons collected by the objective per unit time, once taking into consideration the losses in the collection path and the sensitivity of the detectors (correction factor of 550). (d) Statistics of the collected emission intensity for quantum dots in bulk (outside the micro-pillars, about 50 lines) and within micro-pillars (about 100 lines), gray and red histograms respectively. The solid lines are log-normal fits to the histograms. }
\end{figure*}

We then characterise the emission properties of the quantum dots by means of micro-photoluminescence spectroscopy (a schematic of the set-up can be found in \cite{back_refl})
: we observe a clear enhancement in the collected intensity of the quantum dot emission lines, as shown in Fig.\,3(a-c), when the emitters are located within the pillars, compared to the case where only the back reflector is present. This is due to the channelling of part of the light that would otherwise propagate in the plane towards the collection optics above the pillars, as shown in the simulations on Fig.\,2. Figure\,3d shows the statistics acquired by measuring the collected intensity of dot lines in bulk and from emitters located within the pillars: the histograms are fitted with log-normal distributions that peak at 7.5\,$\times$\,10$^4$  and 23\,$\times$\,10$^4$  photons per second, respectively, and count rates as high as 8\,$\times$\,10$^5$ photons per second are recorded. These results show how a simple pillar/back reflector geometry, made possible by the implementation of the transfer technique that we have developed, can increase the collection efficiency from single quantum dot lines. Other device geometries can be realised using this technique, for instance circular Bragg grating cavities with a gold back reflector \cite{Jin_Au}, planar antennas \cite{Brian}, plasmonic structures \cite{plasmonics}, in an easy and repeatable manner.

In conclusion, we have shown how a technique based on thermal release adhesive tape can be used to pick up and deposit membranes onto host substrates, with an easy and versatile process that is transferred- and host-material independent. We have used this approach to transfer GaAs membranes containing InAs quantum dots onto a metal back reflector layer and we have shown how, combining the broadband reflection of gold with the fabrication on micro-pillars, we can enhance the collection efficiency of quantum light out of the chip. We foresee that this approach will allow the realisation of hybrid devices for classical and quantum photonics, taking advantage of the possibility of stacking relatively large areas of different materials and carrying out nano-fabrication processes on the final layered stack (or on each layer after transfer), without requiring very accurate pick-and-place processes of nano-photonic devices.

\section*{Acknowledgments}

LS acknowledges financial support by the Leverhulme Trust, grant IAF-2019-013.

\newpage


\begin{thebibliography}{50}
\bibitem{disc} Certain commercial products or company names are identified in this paper to foster understanding. Such identification is not intended to imply recommendation or endorsement by the National Institute of Standards and Technology, nor is it intended to imply that the products or names identified are necessarily the best available for the purpose.
\end{thebibliography}
\end{document}